\documentclass[twocolumn,showpacs,preprintnumbers,amsmath,amssymb,prb,tightenlines,final]{revtex4}
\usepackage{amsfonts,amssymb,amsmath,times}
\usepackage{hyperref,graphics} 

\usepackage{graphicx}
\usepackage{dcolumn}
\usepackage{bm}
\usepackage{epsfig}
\usepackage{longtable}

\input epsf

\begin{document}

\title{Effects of self-interaction corrections on the transport properties of phenyl-based molecular junctions} 
\date{\today} 
\author{C. Toher and S. Sanvito}
\email{sanvitos@tcd.ie}
\affiliation{School of Physics and CRANN, Trinity College, Dublin 2, Ireland}

\begin{abstract}
In transport calculations for molecular junctions based on density functional theory the choice of exchange and 
correlation functional may dramatically affect the results. In particular local and semi-local functionals tend to 
over-delocalize the molecular levels thus artificially increasing their broadening. In addition the same
molecular levels are usually misplaced with respect to the Fermi level of the electrodes. These shortfalls 
are reminiscent of the inability of local functionals to describe Mott-Hubbard insulators, but they can
be corrected with a simple and computationally undemanding self-interaction correction scheme.
We apply such a scheme, as implemented in our transport code {\it Smeagol}, to a variety of
phenyl-based molecular junctions attached to gold electrodes. In general the corrections
reduce the current, since the resonant Kohn-Sham states of the molecule are shifted away from
the contact Fermi level. In contrast, when the junction is already described as insulating by local
exchange and correlation potentials, the corrections are minimal and the $I$-$V$ is only
weakly modified.
\end{abstract}


\maketitle

\section{Introduction}

Devices in which organic molecules form the active element are taking an increasingly important place in 
solid state physics. These are the building blocks of the field of molecular electronics, whose potential applications underpin 
a vast number of technological areas. Novel components for high-performance computers \cite{gates1}, highly 
sensitive chemical sensors \cite{chem}, and disposable and wearable devices are among the many proposed applications. 
In addition the same set-up can be used for medical purposes, for instance in the detection of 
viruses \cite{virus} or in the construction of a rapid and reliable protocol for DNA sequencing \cite{MDV-DNA}.

Most of the progress originates from the recently achieved ability to construct single molecule 
junctions and to measure their transport properties. The experimental strategies for fabricating such devices 
include mechanically controllable break junctions \cite{Reed,Tsutsui,Ulrich}, scanning tunneling spectroscopy 
\cite{Ulrich,Tao}, lithographically fabricated nanoelectrodes \cite{Ghosh} and colloid solutions \cite{Dadosh}. 
Unfortunately most of these techniques are actually ``blind'' in the sense that the geometrical configuration
of the device at the atomic scale is not known. For this reason it should not be surprising that different 
experiments for the same molecule yield conductances differing by of up to three orders of magnitude 
\cite{Reed,Tsutsui,Tao,Dadosh}. In view of all these uncertainties {\it ab initio} quantum transport schemes 
for calculating the electrical response of a molecular device become crucial.

The most common computational scheme for evaluating the electronic transport of a molecular device
combines scattering theory in the form of the non-equilibrium Green's function (NEGF) formalism \cite{negf}, 
with an electronic structure method, the most widely used being density-functional theory (DFT) \cite{dft,smeagol,negf-dft}.
Alternatives are time-dependent DFT \cite{tddft} or many-body methods \cite{MB}, which at present however are 
more computationally involved and are at an earlier stage of development as routinely used schemes. 
Unfortunately for several molecules NEGF combined with DFT appears to disagree with the
experimental results. In particular it seems to systematically overestimate the current flowing across a device,
even if one bears in mind the uncertainty over the detailed geometry of the junction. In the prototypical 
case of benzenedithiol (BDT) sandwiched between two gold electrodes, the conductance for the most probable 
contact geometry calculated with NEGF and DFT is higher than that of {\it any} of the experiments by at least one 
order of magnitude \cite{DiVentra,Ratner_1,Ratner_2,Ratner_3}. 

In this work we demonstrate that most of the errors originate from using local or semi-local exchange and
correlation potentials in DFT. This shortfall however can be corrected by applying an atomic self-interaction 
correction (ASIC)\cite{dasc} scheme to the local functionals, leading to a dramatic improvement of the agreement between 
the calculated results and those obtained from experiments. The reason for this improvement is rooted in the ability
of ASIC to predict the correct ionization potential of molecules, and hence to correctly describe the band alignment between 
the molecular orbitals and the Fermi level, $E_\mathrm{F}$, of the metallic electrodes. Here we extend the results previously 
published for BDT attached to the gold hollow site of the (111) surface \cite{bdtsic} to new molecules such as 
benzenedimethanethiol (BDMT) and biphenyldithiol (BPD). We also carry out a thorough investigation 
of the effects of changing the anchoring geometry between the molecule and the electrodes, demonstrating 
that, for the case of BDT attached to gold, the ASIC calculated $I$-$V$ curves are rather stable with respect to geometry changes.
This explains the relatively narrow distribution of conductance histograms found in experiments \cite{Tao}.

\section{Method}

\subsection{Non-Equilibrium Green's Function Formalism}

\begin{figure}[ht]
\begin{center}
\includegraphics[width=6.5cm,clip=true]{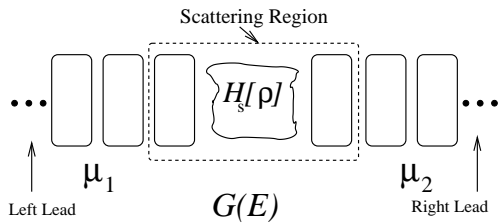}
\end{center}
\caption{\small{Schematic diagram of a metal-molecule-metal junction. A scattering region 
is sandwiched between two current/voltage probes kept at the chemical potentials $\mu_\mathrm{1}$ 
and $\mu_\mathrm{2}$. The electrodes are modeled as being periodic in the direction of transport. 
A number of layers of the electrodes are included in the scattering region (the part of 
the system enclosed by the dashed box) to allow the charge density to converge to the bulk 
value.}}
\label{Fig1}
\end{figure}

The problem we wish to solve is that of calculating the two-probe $I$-$V$ curve of a molecule
sandwiched in between two metallic electrodes. The NEGF scheme partitions such system into
three regions, respectively the two current/voltage electrodes (leads) and a middle region called 
the scattering region (SR) (see figure \ref{Fig1}).

The SR includes the molecule and a portion of the leads, which is extended enough for the
charge density calculated at the most external atomic layer to resemble that of the bulk electrodes. 
The leads, assumed to be a periodic crystal in the direction of the transport, are kept at different chemical potentials 
$\mu_\mathrm{1/2}=E_\mathrm{F}\pm eV/2$ where $V$ is the applied potential bias and $e$ is the 
electron charge. The SR is described by a Hamiltonian $H_\mathrm{s}$, which is used to construct the 
non-equilibrium Green's function, $G(E)$

\begin{equation}
G(E)=\lim_{\eta\rightarrow 0}[(E+i\eta)-H_\mathrm{s}-\Sigma_\mathrm{1}-\Sigma_\mathrm{2}]^{-1}\:,
\label{negfmx}
\end{equation}
where $\Sigma_\mathrm{1/2}$ are the self-energies for the leads. $G(E)$ enters in a self-consistent
procedure to calculate the density matrix, $\rho$, of the SR, and hence the two probe-current of the 
device \cite{negf,smeagol,negf-dft,negftb}. 
The non-equilibrium $\rho$ is calculated following the NEGF prescription as
\begin{equation}
\rho = \frac{1}{2 \pi} \int^{\infty}_{-\infty} G(E)[\Gamma_\mathrm{1} f(E, \mu_\mathrm{1}) +  \Gamma_\mathrm{2} f(E, \mu_\mathrm{2})] G^{\dag}(E) dE \:,
\label{chdens}
\end{equation}
where $\Gamma_\mathrm{1/2} = i[\Sigma_\mathrm{1/2} - \Sigma^{\dag}_\mathrm{1/2}]$.
In practice, this integral is performed by splitting it into two parts\cite{negf,smeagol,negf-dft,negftb}: an equilibrium 
part which can be integrated along a contour in the complex plane, and a non-equilibrium part which has to be 
integrated along the real energy axis but which only contributes around $E_\mathrm{F}$. The non-equilibrium charge density 
$\rho$ is then used to calculate a new Hamiltonian for the scattering region $H_\mathrm{s}[\rho]$ (where it is assumed that the Hamiltonian 
has some functional dependence on $\rho$). This procedure is repeated self-consistently until the density matrix 
converges. Finally, the converged Green's function can be used to calculate the current $I$ through the device:
\begin{equation}
I = \frac{2e}{h} \int^{\infty}_{-\infty} Tr[G(E )\Gamma_\mathrm{1} G^{\dag}(E) \Gamma_\mathrm{2}] (f(E, \mu_\mathrm{1}) -  f(E, \mu_\mathrm{2})) dE \:.
\label{curint}
\end{equation}
This is effectively the integral between $\mu_\mathrm{1}$ and $\mu_\mathrm{2}$ (the bias window) 
of the transmission coefficients $T(E) = Tr[G(E )\Gamma_\mathrm{1} G^{\dag}(E) \Gamma_\mathrm{2}]$.
In brief the NEGF scheme calculates the non-equilibrium scattering potential of the device. Thus the 
transmission coefficient $T(E)$ is simply a superposition of resonances located at the molecular 
single-particle energy levels, including a possible shift and broadening due to the interaction
with the leads. It follows that the molecular levels close to the electrode Fermi level, i.e. the highest
highest occupied molecular orbital (HOMO) and the lowest unoccupied molecular orbital (LUMO),
provide the dominant contribution to the current. It is therefore crucial to employ an electronic
structure theory capable of describing the position of these two levels accurately. 
\begin{figure}[ht]
\begin{center}
\includegraphics[width=6.5cm,clip=true]{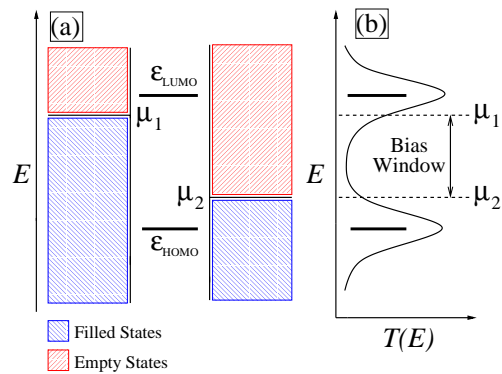}
\end{center}
\caption{\small{Energy level diagram of the metal-molecule-metal junction shown in figure \ref{Fig1}. (a) Energy 
level line up and (b) transmission coefficients as a function of energy. The resonances in the transmission 
coefficients correspond to energy levels in the molecule.}}
\label{Fig2}
\end{figure}

\subsection{DFT and the self-interaction problem}

The NEGF scheme is general and not related to a particular electronic structure
theory. However, since the transmission occurs through single-particle states, the associated electronic structure theory 
should meet several requirements \cite{tbsic}. First of all, the single-particle levels must closely 
resemble the physical removal energies of the system, i.e. the energy levels of the molecule should line up correctly with 
the $E_\mathrm{F}$ of the metals forming the electrodes.  Secondly, the theory should work at both integer and fractional 
occupation, ensuring the correct response of the molecular orbitals to changes in occupation due to 
the applied potential bias. Finally, the matrix elements describing the interaction between the leads and the molecule 
should be calculated accurately avoiding the use of phenomenological parameters or fitting procedures.

The Kohn-Sham (KS) form \cite{ks} of DFT is certainly the most widely used {\it ab initio} electronic structure theory 
associated with the NEGF method. In the KS scheme the DFT problem of finding the ground state charge density is mapped 
onto that of solving a system of non-interacting single-particle Schr{\" o}dinger equations
\begin{equation}
[-\frac{1}{2} \nabla^2 + v({\bf r}) + u(\rho) + v_\mathrm{XC} (\rho, {\bf r})] \psi^{\sigma}_n({\bf r}) = \epsilon^{\sigma}_n \psi^{\sigma}_n({\bf r}) \:,
\label{kseqn}
\end{equation}
where $u(\rho)$ is the Hartree (classical Coulomb) potential and $v_\mathrm{XC}$ is the exchange-correlation (XC) potential. 
The exact form of $v_\mathrm{XC}$ is unknown, so it is typically approximated by local functionals such as the local density 
approximation \cite{ks} (LDA) or the generalized gradient approximation \cite{PBE} (GGA). 

Therefore the KS scheme is only a convenient prescription for minimizing the energy functional, and as such the individual KS 
eigenvalues do not necessarily correspond to physical energy levels of the system. The DFT observables such as the total energy
and the charge density are in fact only integral quantities respectively of the KS eigenvalues and eigenvectors. However, in solid state 
theory it is common to associate the KS energies with the system band structure. This, although it is not justified at the fundamental 
level, is supported by the practical evidence that KS bands represent a good first approximation of the true energies of a system, 
particularly for metals. However, there is a remarkable exception to the non-physical nature of the KS eigenvalues. In fact, the KS energy of the HOMO level 
($\epsilon^\mathrm{KS}_\mathrm{HOMO}$) can be rigorously associated with the negative of the ionization potential $I$ \cite{janak,PPLB}. 
Similarly the HOMO KS energy for the negatively charged system can be associated with the chemical affinity $A$.
This suggests that, at least for moderate bias where the transport is through the HOMO, KS theory can be used effectively for 
transport calculations \cite{note}. Unfortunately, standard local and semi-local functionals completely misplace
$\epsilon^\mathrm{KS}_\mathrm{HOMO}$, which is often nowhere near $-I$. This is particularly problematic for 
organic molecules, for which the LDA $\epsilon^\mathrm{KS}_\mathrm{HOMO}$ is typically off from the experimental value of $I$ by 
approximately 4~eV \cite{dasc}.

Most of the failures of LDA and GGA can be traced back to the self-interaction (SI) problem, i.e. the spurious interaction of an 
electron with the Hartree and exchange-correlation (XC) potentials generated by itself \cite{PZ}. In the case of Hartree-Fock theory 
the self-Hartree energy $U[\rho^{\sigma}_n]$ cancels exactly the self-XC energy $E_\mathrm{XC}[\rho^{\sigma}_n, 0]$
\begin{equation}
E_\mathrm{XC}[\rho^{\sigma}_n, 0] + U[\rho^{\sigma}_n] = 0 \:,
\label{exxhar}
\end{equation}
where we introduce the orbital density $\rho^{\sigma}_n = |\psi^{\sigma}_n|^2$.
However, this cancellation is incomplete for both LDA and GGA. The resulting KS potential therefore appears too repulsive 
and the eigenstates of a molecule are pushed to higher energies. When this is transferred to the transport problem, the position 
of the peaks in the transmission coefficient results are in the wrong place. In particular it is likely that peaks will erroneously move close
to the Fermi level of the electrodes, generating a large conductivity at zero bias. Since the position of the peaks is ultimately
determined by the interplay between the hopping probabilities from the electrodes to the molecule with the charging energy
of the molecule itself, this failure is somewhat reminiscent of the inability of local and semi-local functionals to describe 
Mott-Hubbard insulators. 

Perdew and Zunger suggested \cite{PZ} that the self-interaction correction (SIC) $\delta^\mathrm{SIC}_n$ for an occupied KS 
orbital $\psi^{\sigma}_n$ can be calculated from the sum of the self-Hartree and self-XC energies
\begin{equation}
\delta^\mathrm{SIC}_n = E^\mathrm{LDA}_\mathrm{XC}[\rho^{\sigma}_n, 0] + U[\rho^{\sigma}_n] \:.
\label{siclda}
\end{equation}
Note that, although here we consider
the case of SIC-LDA only, the same procedure can be readily applied to any approximated functionals. The SIC-LDA XC 
energy $E^\mathrm{SIC}_\mathrm{XC}$ is thus obtained by subtraction 
\begin{equation}
E^\mathrm{SIC}_\mathrm{XC}[\rho^{\uparrow}, \rho^{\downarrow}] = E^\mathrm{LDA}_\mathrm{XC}[\rho^{\uparrow}, \rho^{\downarrow}] - 
\sum^\mathrm{occ.}_{n \sigma} \delta^\mathrm{SIC}_n  \:,
\label{sicen}
\end{equation}
and the SIC-LDA XC potential $v^{\mathrm{SIC},\sigma}_n$, to be subtracted from $v_\mathrm{XC}^{\mathrm{LDA},\sigma}$, 
is simply given by
\begin{eqnarray}
v^{\mathrm{SIC},\sigma}_n=u([\rho_n]; {\bf r})+v_\mathrm{XC}^{\mathrm{LDA},\sigma}([\rho_n^\uparrow,0]; {\bf r})
\;,
\label{sicpot}
\end{eqnarray}
with 
\begin{eqnarray}
u([\rho]; {\bf r})=\int\mathrm{d}^3{\bf r}^\prime\frac{\rho({\bf r}^\prime)}{|{\bf r}-{\bf r}^\prime|}\;, \\
v_\mathrm{XC}^{\mathrm{LDA},\sigma}([\rho^\uparrow,\rho^\downarrow]; {\bf r})=\frac{\delta}{\delta \rho^\sigma({\bf r})}
E_\mathrm{XC}^\mathrm{LDA}[\rho^\uparrow,\rho^\downarrow]\;.
\end{eqnarray}

The problem of finding the energy minimum is complicated by the fact that $E^\mathrm{SIC}_\mathrm{XC}$ is not invariant 
under unitary rotations of the occupied KS orbitals, which instead leave $\rho$ invariant. This creates problem to the standard
KS scheme since the theory becomes size-inconsistent. In order to avoid such a complication modern SIC theory introduces
a second set of orbitals $\phi^{\sigma}_n$  related to the canonical KS orbitals $\psi^{\sigma}_n$ by a unitary transformation $M$
\begin{equation}
\psi^{\sigma}_n = \sum_m M^{\sigma}_{nm} \phi^{\sigma}_n \:.
\label{uniorb}
\end{equation}
The functional can then be minimized by varying both the KS orbitals and the unitary transformation $M$, 
leading to the system of KS-like equations: 
\begin{equation}
H^{\sigma}_n \psi^{\sigma}_n = [H^{\sigma}_0 + \Delta v^\mathrm{SIC}_n] \psi^{\sigma}_n({\bf r}) = \epsilon^{\sigma, \mathrm{SIC}}_n \psi^{\sigma}_n({\bf r}) \:,
\label{siceqn}
\end{equation}
where $H^{\sigma}_0$ is the LDA Hamiltonian, and the SIC potential, $\Delta v^\mathrm{SIC}_n$, is given by:
\begin{equation}
\Delta v^\mathrm{SIC}_n = \sum_m M^{\sigma}_{nm} v^\mathrm{SIC}_m \frac{\phi^{\sigma}_n}{\psi^{\sigma}_n} =  \sum_m v^\mathrm{SIC}_m \hat{P}^{\phi}_m \:,
\label{dvsic}
\end{equation}
where $\hat{P}^{\phi}_m$ is the projector $|\phi^{\sigma}_m \rangle \langle \phi^{\sigma}_m|$. 

The numerical implementation of the full Perdew-Zunger SIC scheme in typical solid state codes is 
cumbersome since the theory is orbital dependent and the energy minimization cannot follow the 
standard KS scheme \cite{dasc}. When this is applied to transport there is the additional complication
that the KS orbitals are never individually available and moreover that one has to deal with an intrinsically
non-local potential. A drastic simplification of the problem can be obtained by using a recently implemented 
atomic approximation to the SIC scheme, which we call the ASIC \cite{dasc}. This is based on the 
pseudo-SIC approximation, originally proposed by Vogel and coworkers \cite{vogel}, and later
extended to non-integer occupation by Filippetti and Spaldin \cite{pseudoSIC}. The main idea is
that of replacing the $\phi_m$ orbitals with atomic like functions, which are not calculated
self-consistently. Thus the SI correction becomes atomic-like and no information is needed other 
than the charge density and the ASIC projectors \cite{dasc}. 

ASIC has been demonstrated to produce single-particle energy levels which match the experimental 
molecular removal energies quite well \cite{dasc}. In particular, for several different molecules investigated,  
$\epsilon^\mathrm{KS}_\mathrm{HOMO}\approx -I$, and the $\epsilon^\mathrm{KS}_\mathrm{HOMO}$ for negatively charged molecules is 
close to the molecular affinity \cite{dasc}. As an example, ASIC places $\epsilon^\mathrm{KS}_\mathrm{HOMO}$ for 
1,2-BDT at -8.47~eV to compare with the LDA value of -4.89~eV and the experimental ionization potential $\sim$8.5~eV \cite{shen}.
ASIC is the scheme that we adopt in this work.

\subsection{ASIC and the energy derivative discontinuity} 
\label{enderdis}

Local and semi-local functionals are affected by another fundamental problem, i.e. the lack of the derivative discontinuity
(DD) in the DFT energy. This is the discontinuity at integer occupation in the derivative of the total energy of a system $E(N)$
with respect to its occupation $N$ \cite{PPLB}. From Janak's theorem it follows that there should be also a discontinuity 
in $\epsilon^\mathrm{KS}_\mathrm{HOMO}(N)$ when going from $N$ to ($N+\delta$) with $\delta\rightarrow 0^+$. 

In a previous work \cite{tbsic}, we have investigated the consequences of the absence of the DD in local XC functionals on the 
electronic transport of organic molecules. We have demonstrated that the DD can strongly affect the $I$-$V$ characteristics of 
metal-molecule-metal junctions when the coupling between the molecule and the metal is relatively weak. This would be 
the case, for instance, where molecules bind to adatoms on the metal surface. However, we have also shown that in the case of strong 
coupling the DD had little effect on the transport.

The self-interaction error is largely responsible for the disappearance of the DD in local and semi-local functionals.
It follows that SIC methods restore the DD at least in part. This last aspect is deeply rooted in the orbital minimization needed
to extract the SIC potential, which unfortunately is not captured by our ASIC approximation. Therefore the main feature of
our ASIC methodology is simply the correction of the ionization potentials of the molecules \cite{dasc}, thus yielding a 
quantitatively realistic energy level alignment. Hence, the calculations described in this paper investigate a second important 
aspect of the SIC beside that of the DD.  This is the quantitative description of the band alignment between the metal Fermi energy 
and the molecular levels. In particular, for most of the paper we investigate molecules attached to the gold fcc (111) hollow sites, 
for which the coupling is expected to be strong and the effects due to the DD small.

\subsection{Calculation Details}

We have numerically implemented the ASIC method \cite{dasc} in the localized atomic orbital code SIESTA \cite{siesta},
which is the DFT platform for our transport code {\it Smeagol} \cite{smeagol}, and carried out calculations for the 
prototypical Au/Benzene/Au molecular devices. The ASIC corrections are not applied to the gold atoms, as SI errors for
metals are small \cite{dasc}. Unless otherwise specified, we use a double zeta polarized basis set \cite{siesta} for carbon 
and sulfur $s$ and $p$ orbitals, double zeta for the 1$s$ orbitals of hydrogen and 6$s$-only double zeta for gold. The 
mesh cut-off is 200 Ry and we consider 500 real and 80 complex energy points for integrating the Green's function. Calculations 
were also performed using double zeta 6$s$ and single zeta 5$d$ and 6$p$ orbitals for the gold to investigate the effect of an
enlarged basis on the transport. Results for calculations using both basis sets are shown, with the 6$s$-only basis used 
unless otherwise indicated. For geometry optimizations and total energy calculations, the extended 5$d$6$s$6$p$ basis set is 
employed, as the 6$s$-only does not yield enough accuracy. In calculating the $I$-$V$ curves, the potential bias was incremented 
in steps of 0.1 Volts.

\section{Results} 

We compare the electronic transport properties for three different molecules calculated either by using LDA or ASIC. These 
are 1,4-benzenedithiol (1,4-BDT), benzenedimethanethiol (BDMT), and biphenyldithiol (BPD). The electronic transport through 
these molecules have already been investigated at length both experimentally and computationally \cite{Reed,Tsutsui,Ulrich,
Tao,Dadosh,Ghosh,MB,DiVentra,Ratner_1,Ratner_2,Ratner_3}. Thus these offer us a unique benchmark for our calculations.
In all cases, the molecule is attached via the sulfur atoms to fcc (111) gold electrodes on each side. In the case of BDT, we look 
at several different anchoring geometries. It is  worth pointing out that although the exact anchoring geometry encountered in 
breaking junction experiments is unknown, the resulting conductance histograms are relatively narrow suggesting an
intrinsic stability in the measurements \cite{Tao}. This may be an indication that similar anchoring configuration are highly probable in
the breaking process, or alternatively that several anchoring geometries yield the same $I$-$V$ characteristic.
One of our goals is to distinguish between these two possibilities. 

\subsection{Benzenedithiol}
 
\begin{figure}[ht]
\begin{center}
\includegraphics[width=8.0cm,clip=true]{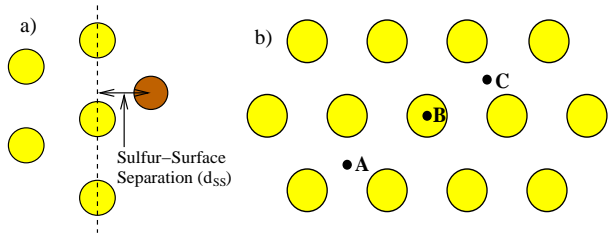}
\end{center}
\caption{\small{(Color on line). The different anchoring geometries investigated. Panel (a) defines the ``sulfur-surface separation'', $d_\mathrm{SS}$, as
the distance between the S anchoring ion and the Au fcc (111) plane. In (b) we show the different possible anchoring sites.
{\bf A} is the hollow site, {\bf B} is the top site, and {\bf C} is the bridge site. Color code: Au atoms=yellow (or light gray), 
S atoms=brown (or dark gray).}}
\label{Fig3}
\end{figure}

The first device which we consider is 1,4-benzenedithiol (1,4-BDT) attached to two fcc (111) gold surfaces. 
The anchoring geometries for the sulfur atoms investigated are the hollow site, the top site, the bridge site, 
(see figure \ref{Fig3}), as well as asymmetric configurations where the S ions are attached to an adatom on 
one side and to the hollow site on the other. We also examine the effect of altering the angle which the molecule makes with 
the metal surface, and of varying the distance between the sulfur atom and the surface (i.e. varying the strength of the 
coupling between the molecule and the metal). Finally, we investigate the effect of hydrogenating the thiol groups.

As explained previously the actual experimental contact geometry is unknown. Electronic structure calculations indicate 
that the lowest energy configuration for the molecule on the (111) surface occurs when the S atom attaches to the hollow 
site, although other calculations suggest that the bridge site configuration has a lower energy \cite{Bridge}. Recent X-ray standing 
wave measurements suggest that molecules in monolayers prefer to attach to adatoms on the metal surface \cite{adatom_thiol}. 
It is also worth noticing that in breaking junctions the actual geometry might be different from any equilibrium geometries,
since the system is likely to be under strain. Hence, a thorough analysis of this system should comprise many different configurations.

\begin{figure}[ht]
\begin{center}
\includegraphics[width=7.5cm,clip=true]{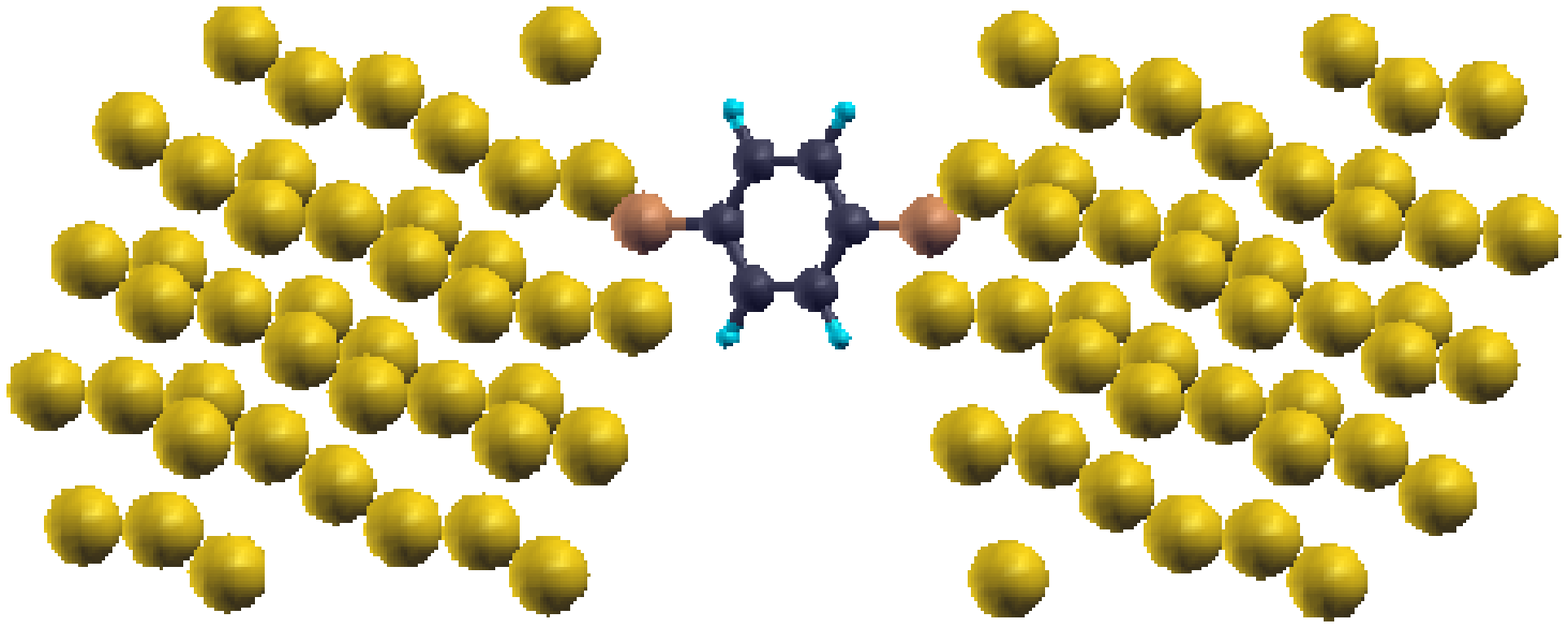}
\end{center}
\caption{\small{(Color on line). BDT molecule attached to the hollow site of the Au (111) surface. 
The sulfur-surface distance, $d_\mathrm{SS}$, is 1.9~\AA. Color code: Au=yellow (or light gray), C=black, 
S=dark yellow (or gray), H=blue (or dark gray).}}
\label{Fig4}
\end{figure}

We start by attaching BDT at the Au (111) hollow site (see figures \ref{Fig3} and \ref{Fig4}). The distance of the sulfur atom from 
the plane of the gold surface (the ``sulfur-surface separation'', $d_\mathrm{SS}$, shown in figure \ref{Fig3}a) is optimized (using the 5$d$6$s$6$p$ 
basis set for gold) to a value of 1.9~\AA. This corresponds to a distance of 2.53~\AA ~between the sulfur atoms and the nearest gold 
atoms on the surface in good agreement with previous calculations \cite{Ratner_1, AuSdist, Bridge}.

\begin{figure}[ht]
\begin{center}
\includegraphics[width=9.0cm,clip=true]{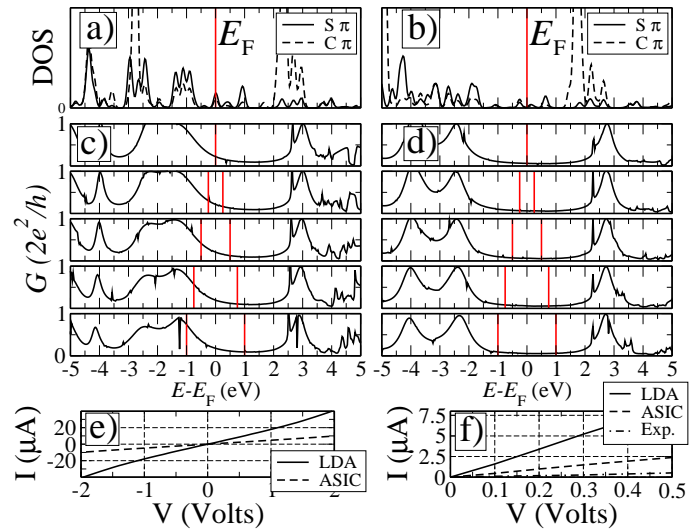}
\end{center}
\caption{\small{Transport properties of a BDT molecule attached to the gold fcc (111) hollow site. The left
plots correspond to LDA and the right ones to ASIC. The upper panels are the DOS of the S 
and C $\pi$ orbitals ((a) and (b)), the middle are the transmission coefficients as a function 
of energy for various bias ((c) and (d)) and the lower are the $I$-$V$ curves. Figure (f) is a zoom of
(e) and compares our results with experiments from reference \cite{Tao}. The vertical lines in
(c) and (d) mark the bias window.}}
\label{Fig5}
\end{figure}

The $I$-$V$ characteristic, the orbital resolved density of states (DOS) and the conductance as a function of energy,
$G$, are presented in figure \ref{Fig5}. The conductance is plotted in units of $\frac{2e^2}{h}$, so that it is equivalent 
to the transmission coefficients $T$. LDA results are shown in the left panels, while the ASIC ones are on the right.
From the DOS it is clear that the effect of ASIC is that of shifting the occupied orbitals downwards in energy
relatively to the $E_\mathrm{F}$ of gold. The ASIC HOMO-LUMO gap is considerably larger than that calculated with LDA
and most importantly in the case of ASIC there is little DOS originating from the molecule at $E_\mathrm{F}$.
This has profound effects on the electron transmission. The LDA peaks of $T(E)$
arising from occupied orbitals are shifted downwards in energy and away from $E_\mathrm{F}$. 
At variance from LDA (figure \ref{Fig5}c), where $T(E_\mathrm{F})$ is dominated 
by a resonance at $\epsilon_\mathrm{HOMO}$, the ASIC transmission (figure \ref{Fig5}d) is through the BDT gap
and therefore it is tunneling-like. This results in a drastic reduction of the low-bias current when going from
LDA to ASIC (figure \ref{Fig5}e). The ASIC-calculated conductance at zero bias is now about 0.06$G_0$ ($G_0=2e^2/h$), 
compared to 0.23$G_0$ for LDA. A conductance of 0.06$G_0$ is much closer to the value of 
0.011$G_0$ obtained by Xiao et. al. \cite{Tao} and is actually lower than values 0.09-0.14$G_0$ 
obtained by Tsutsui et. al. \cite{Tsutsui}.

\begin{figure}[ht]
\begin{center}
\includegraphics[width=9.0cm,clip=true]{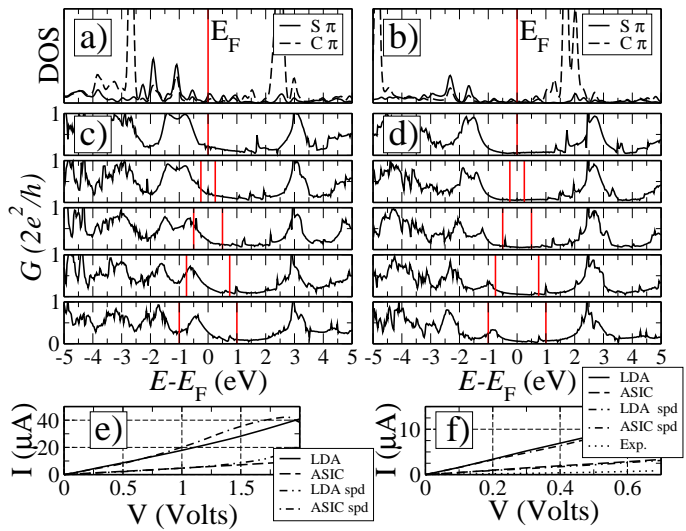}
\end{center}
\caption{\small{Transport properties of a BDT molecule attached to the Au fcc (111) hollow site calculated with
an enriched 5$d$6$s$6$p$ basis set for gold. The left plots correspond to LDA and the right ones to ASIC results. 
The upper panels are the DOS of the S and C $\pi$ orbitals ((a) and (b)), the middle are the transmission 
coefficients as a function of energy for various bias ((c) and (d)) and the lower ((e) and (f)) are the $I$-$V$ 
curves, including a comparison with the results shown in figure \ref{Fig4} for the 6$s$-only basis. Figure (f) 
is a zoom of (e) and compares our results with experiments from reference \cite{Tao}. 
The vertical lines in (c) and (d) mark the bias window.}}
\label{Fig6}
\end{figure}

Without altering the anchoring geometry, the basis set on the gold atoms is then changed to include the 5$d$ and 6$p$ orbitals. 
The relative orbital resolved DOS, transmission coefficients and $I$-$V$ curves are presented in figure \ref{Fig6} for 
both LDA and ASIC. Clearly the enriched basis set does not have a large effect on the electronic
transport, particularly at low bias. As can be seen from figure \ref{Fig6}e, the $I$-$V$ curves calculated with the 6$s$-only basis 
set for gold are approximately the same as that calculated with the 5$d$6$s$6$p$ basis set up to about 1 Volt for both 
the LDA and ASIC. Differences appear only at high bias ($V>1$~Volt) and are due to the presence of Au $d$ electrons
in an energy range comprising the bias window.
However the zero-bias conductances, from the transmission plots in panels (c) and (d) of figure \ref{Fig6}, 
are very similar to those for the 6$s$-only basis set with a value of 0.21$G_0$ for LDA and a value of 0.06$G_0$ for ASIC. 
This demonstrates that the 6$s$-only basis set should give reasonably reliable results for electronic transport at low bias. 

\begin{figure}[ht]
\begin{center}
\includegraphics[width=9.0cm,clip=true]{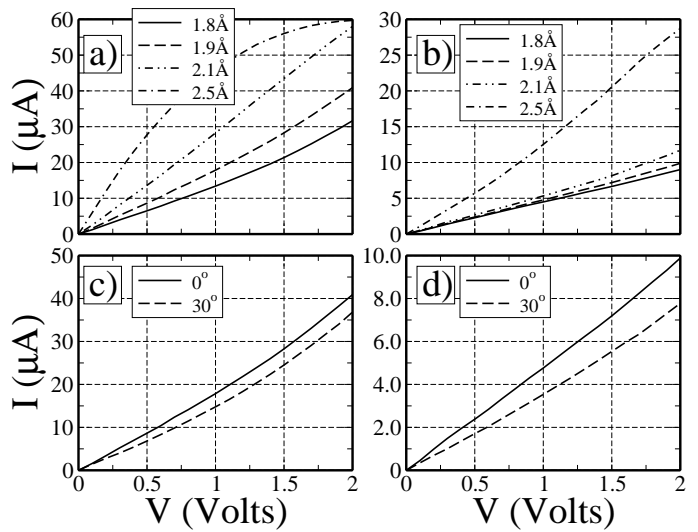}
\end{center}
\caption{\small{$I$-$V$ curves for BDT attached to Au contacts for different $d_\mathrm{SS}$: (a) LDA and (b) ASIC results. 
In (c) and (d) we present LDA and ASIC calculated $I$-$V$ curves where $d_\mathrm{SS}=1.9$~\AA\ but the angle between
the molecule and the direction of the transport is 30$^o$. }}
\label{Fig7}
\end{figure}

Next, the transport properties of the system are calculated for the hollow site at different $d_\mathrm{SS}$,
as well as for the equilibrium distance $d_\mathrm{SS}=1.9$~\AA\ but different angles of the molecule with 
respect to the direction of transport. The results are reported in figure \ref{Fig7}. In general ASIC seems to be much less 
sensitive to changes in the anchoring geometry than LDA, particularly for the case of bond stretching. 
For instance, the zero bias conductance values calculated with LDA are 0.16~$G_0$, 0.23~$G_0$, 0.32~$G_0$ 
and 0.77~$G_0$ for $d_\mathrm{SS}$ values of 1.8~\AA, 1.9~\AA, 2.1~\AA ~and 2.5~\AA ~respectively. For the
same $d_\mathrm{SS}$ ASIC returns 0.05~$G_0$, 0.06~$G_0$, 0.07~$G_0$, and 0.14~$G_0$.

\begin{figure}[ht]
\begin{center}
\includegraphics[width=8.0cm,clip=true]{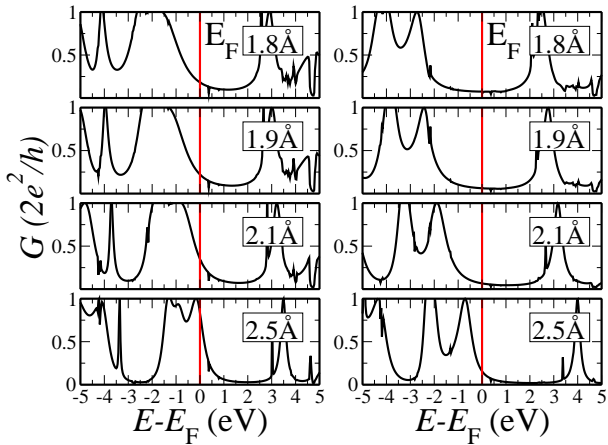}
\end{center}
\caption{\small{Transmission coefficients for BDT attached to Au contacts at different $d_\mathrm{SS}$. The left 
plots correspond to LDA and the right ones to ASIC results. Note how the transmission peaks narrow and how the 
HOMO peak moves closer to the gold $E_\mathrm{F}$ as $d_\mathrm{SS}$ is increased. This realignment of the 
HOMO of the molecule has the effect of compensating for the weakening of the coupling, thus producing the counterintuitive 
result of the low bias conductance increasing by enlarging $d_\mathrm{SS}$.}}
\label{Fig8}
\end{figure}

\begin{figure}[ht]
\begin{center}
\includegraphics[width=9.0cm,clip=true]{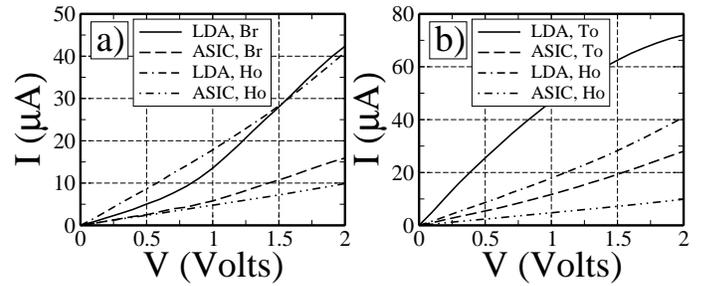}
\end{center}
\caption{\small{$I$-$V$ curves for BDT attached to the Au (111) surfaces via a) bridge site and b) top site. For comparison in both 
cases the $I$-$V$ curves for the hollow site geometry are also reported.}}
\label{Fig9}
\end{figure}

The stability of the ASIC $I$-$V$ curve with respect the details of the anchoring geometry of the hollow site 
is an interesting feature, since it is suggesting that in breaking junction experiments several different
anchoring configurations may yield similar $I$-$V$ curves. This is consistent with the relatively narrow
peaks measured in the typical conductance histograms \cite{Tao}. LDA does not seem to have this property
and in fact the zero-bias conductance increases quite drastically when $d_\mathrm{SS}$ is increased. 
While counterintuitive, this is consistent with previous results \cite{Ratner_1, Ratner_3}, and it is 
due to the realignment of the HOMO of the molecule to $E_\mathrm{F}$ of gold. Such a feature is investigated
in more detail in figure \ref{Fig8} where we report the zero-bias $T(E)$ for different $d_\mathrm{SS}$. 
When $d_\mathrm{SS}$ is increased, the transmission peaks corresponding 
to molecular orbitals become narrower, as expected, due to the weakened molecule-lead coupling. However, charge 
transfer between the molecule and the metal is also affected, so that the charge on the actual molecule slightly increases
as $d_\mathrm{SS}$ gets larger. This extra charge produces an electrostatic shift of the HOMO level towards the
Au Fermi energy. As a consequence the narrowing of the resonance peak at the HOMO level is compensated 
by the upwards shift of the same peak, resulting in an enhancement of the zero-bias conductance. When ASIC
is used this feature is strongly suppressed. In fact, the ASIC calculated transmission peaks are narrower and
further away from $E_\mathrm{F}$ that their LDA counterparts. Thus, although also for ASIC the molecular HOMO 
realigns with respect to the Au Fermi level, this is is not enough to produce high transmission at reasonable 
$d_\mathrm{SS}$. Note that a similar realignment is expected for weaker binding anchoring geometries which we
will investigate in the remainder of this section (see figures \ref{Fig9}b, \ref{Fig11}, \ref{Fig12} and \ref{Fig14})

The second contact geometry investigated is that where the S atom is connected to the bridge site on the 
fcc (111) Au surface at both of the electrodes (figure \ref{Fig3}b). Total energy DFT calculations suggest that this 
configuration ($d_\mathrm{SS}=2.09$~\AA), has a lower energy than that of the hollow site \cite{Bridge}. 
We calculate a LDA zero bias conductance for the bridge site of 0.1~$G_0$, which is lower than the 
value of 0.23~$G_0$ obtained for the hollow site. This can be seen from the $I$-$V$ curves presented in
figure \ref{Fig9}a. Interestingly this lower conductance found for the bridge site with respect to the hollow site
is a low-bias feature and the two $I$-$V$ curves matches closely for $V>1.5$~Volt. The effect of ASIC 
on the transmission of the bridge site configuration is rather similar to that on the hollow site. Also 
in this case the molecular HOMO-LUMO gap opens and the current gets suppressed. The zero bias conductance 
for the bridge site is calculated to be 0.06~$G_0$, the same value found for the hollow site. Hence, whether the 
molecule is anchored to the hollow site or to the bridge site makes relatively little difference to the low-bias transport 
properties as obtained with ASIC. This further confirms that our ASIC calculations are certainly more compatible than
their LDA counterparts with the relative stability of the experimental conductance histograms. 

The top site of the Au fcc (111) surface (figure \ref{Fig3}b) is the next anchoring site to be investigated. 
The equilibrium distance $d_\mathrm{SS}$ is 2.39~\AA\  and the calculated $I$-$V$ curves
are presented in figure \ref{Fig9}b. This time both LDA and ASIC present $I$-$V$ characteristics with
much higher conductance than that associated with the hollow site. This is particularly dramatic for LDA for which
the zero-bias conductance goes from 0.23~$G_0$ for the hollow to 0.65~$G_0$ for the top site. For ASIC
the increase is only twofold with the zero bias conductance going from 0.06~$G_0$ to 0.12~$G_0$.
Also for this case the general increase in current can be correlated with the larger Au-S bond-length 
of the top site compared to the hollow site. This has the same effect found when analyzing the hollow
site under strain and it is due to the realignment of the HOMO of BDT to $E_\mathrm{F}$ of gold.
It is important to point out that the top site is less energetically favorable than both the bridge and
the hollow sites. Although it can still play a role in breaking junctions, we believe that this may be
only of secondary importance and therefore would not have a considerable influence on the typical 
conduction histograms.

\begin{figure}[ht]
\begin{center}
\includegraphics[width=7.5cm,clip=true]{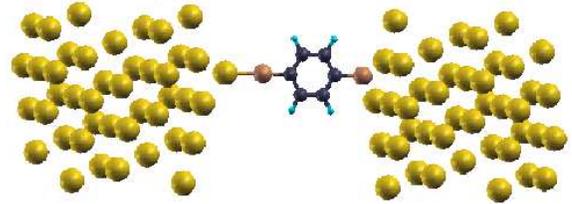}
\end{center}
\caption{\small{(Color on line) BDT molecule attached asymmetrically to two Au (111) surfaces. On one side the bonding
is at a hollow position ($d_\mathrm{SS}=1.9$~\AA), while on the other is to a gold adatom (sulfur-adatom separation=2.39~\AA).
Color code: Au=yellow (light gray), C=black, S=dark yellow (gray), H=blue (dark gray).}}
\label{Fig10}
\end{figure}

\begin{figure}[ht]
\begin{center}
\includegraphics[width=9.0cm,clip=true]{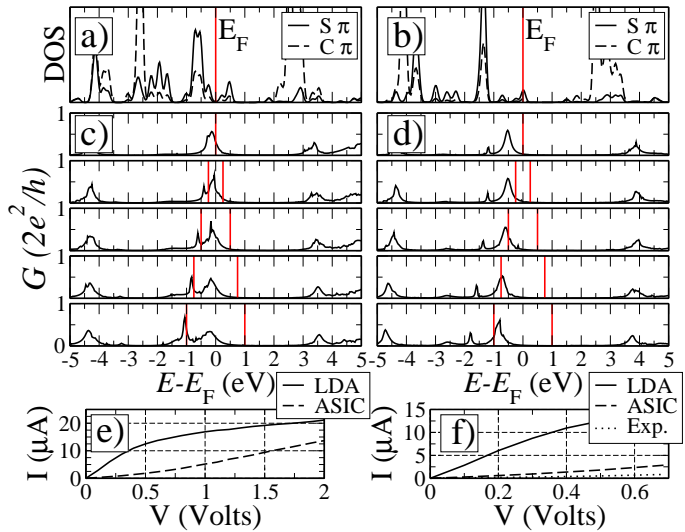}
\end{center}
\caption{\small{Transport properties of a BDT molecule attached asymmetrically to two Au (111) surfaces
according to the anchoring geometry of figure \ref{Fig10}. 
The left plots correspond to LDA and the right ones to ASIC. The upper 
panels are the DOS of the S and C $\pi$ orbitals ((a) and (b)), the middle are the transmission coefficients 
as a function of energy for various bias ((c) and (d)) and the lower are the $I$-$V$ curves. Figure (f) is 
a zoom of (e) and compares our results with experiments from reference \cite{Tao}. The vertical lines in
(c) and (d) mark the bias window.}}
\label{Fig11}
\end{figure}

\begin{figure}[ht]
\begin{center}
\includegraphics[width=9.0cm,clip=true]{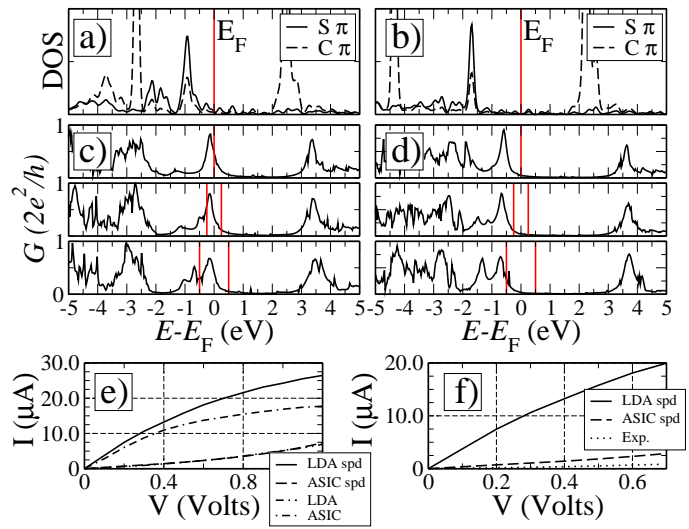}
\end{center}
\caption{\small{Transport properties of a BDT molecule attached asymmetrically to two Au (111) surfaces
according to the anchoring geometry of figure \ref{Fig10}. Results are presented for calculations using 
an enlarged 5$d$6$s$6$p$ basis for Au. The left plots correspond to LDA and the right 
ones to ASIC. The upper panels are the DOS of the S and C $\pi$ orbitals ((a) and (b)), the middle are the transmission 
coefficients as a function of energy for various bias ((c) and (d)) and the lower are the $I$-$V$ curves. Figure (f) 
is a zoom of (e) and compares our results with experiments from reference \cite{Tao}. The vertical lines in (c) and 
(d) mark the bias window.}}
\label{Fig12}
\end{figure}

In actual breaking junctions, the electrodes are under strain and so the thiol groups are less likely to
attach to a regular surface site.
For instance in figure \ref{Fig10} we present the case where
one electrode connects to the molecule via the hollow site ($d_\mathrm{SS}=1.9$~\AA) and the other via
a gold adatom, with a distance of 2.39~\AA\ between the sulfur and the adatom. Recent X-ray standing wave 
experiments \cite{adatom_thiol} demonstrate that gold adatoms may be the most favorable sites for the thiol 
groups in self-assembled molecular monolayer. This therefore may be a likely configuration for the bottom 
electrode in a STM-break junction.

The orbital resolved DOS, transmission coefficients and $I$-$V$ curves for this system are 
presented in figure \ref{Fig11}. This configuration shows the largest difference between the conductance 
calculated with LDA and that calculated with ASIC. The LDA conductance is 0.32~$G_0$, whereas when ASIC is 
applied this drops by a order of magnitude to 0.03~$G_0$. Because of the weaker interaction between the $\pi$ orbitals 
of the molecule and the gold surface, the molecular orbitals in the DOS (figure \ref{Fig11}a and figure \ref{Fig11}b) and hence the 
peaks in the transmission coefficients (figure \ref{Fig11}c and figure \ref{Fig11}d) are considerably narrower
than for the case of the hollow site. These narrow levels are however closer to $E_\mathrm{F}$, resulting in the 
relatively high conductance at low bias. This originates from the same mechanism producing a change in conductance
when $d_\mathrm{SS}$ is increased (see figure \ref{Fig8}). At variance with the case of stretched hollow bonding 
geometry, this time the ASIC does not shift the conductance peaks enough to bring them close to the Fermi level
of gold, and the conductance remains rather small. 

As the peaks in the transmission are rather narrow we checked our results against the choice of basis set. 
In figure \ref{Fig12}  we present the orbital resolved DOS, transmission coefficients and $I$-$V$ 
curves for the enlarged 5$d$6$s$6$p$ basis, to be put in perspective with those calculated with 6$s$ only
in figure \ref{Fig11}. The most notable difference is the appearance of transmission away from $E_\mathrm{F}$,
in particular at low energies. This is connected to the gold $d$ electrons, which are neglected in the small
basis. Importantly however the low-energy features around $E_\mathrm{F}$ are only slightly affected with
the zero-bias conductance assuming values of 0.47~$G_0$ and 0.05$~G_0$ for LDA and ASIC respectively.
This confirms the feasibility of using a restricted basis set for the $I$-$V$ curve at low bias. 

\begin{figure}[ht]
\begin{center}
\includegraphics[width=7.5cm,clip=true]{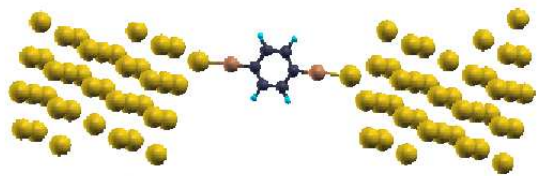}
\end{center}
\caption{\small{(Color on line) BDT molecule attached to adatoms at both of the Au (111) surfaces.   
Color code: Au=yellow (or light gray), C=black, S=dark yellow (or gray), H=blue (or dark gray).}}
\label{Fig13}
\end{figure}

\begin{figure}[ht]
\begin{center}
\includegraphics[width=7.5cm,clip=true]{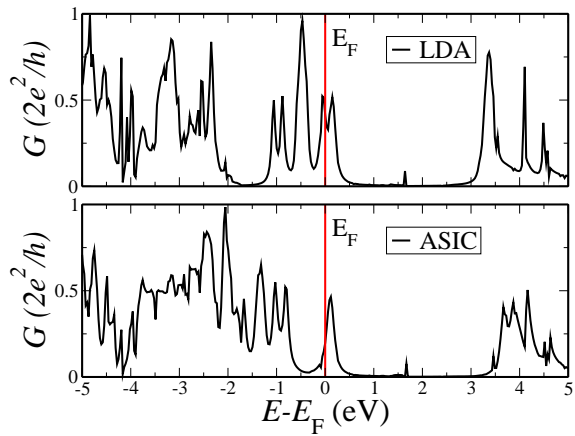}
\end{center}
\caption{\small{Transmission  coefficients for a BDT molecule attached to adatoms on both the gold (111) 
surfaces, calculated with the 5$d$6$s$6$p$ basis for gold. Note that the HOMO is now pinned at the gold $E_\mathrm{F}$.}}
\label{Fig14}
\end{figure}

\begin{figure}[ht]
\begin{center}
\includegraphics[width=7.5cm,clip=true]{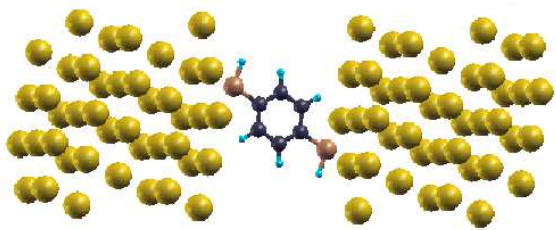}
\end{center}
\caption{\small{(Color on line) BDT molecule with H atoms attached to the S atoms in the thiol groups ($d_\mathrm{SS}=1.9$~\AA). 
Color code: Au=yellow (or light gray), C=black, S=dark yellow (or gray), H=blue (or dark gray).}}
\label{Fig15}
\end{figure}

Finally, we investigate two more configurations. The first considers adatoms as the bonding site at both sides
of the junction (figure \ref{Fig13}), while the second investigates hollow sites where the two hydrogens of the thiol
groups are not dissociated in forming thiolate as in all the other cases studied (figure \ref{Fig15}).
In the first case we expect a rather weak bond, which is confirmed by the sharp peaks in transmission
shown in figure \ref{Fig14}. The LDA low-bias conductance, calculated with the 5$d$6$s$6$p$ basis for gold, 
is 0.43~$G_0$, which drops to 0.19~$G_0$ when ASIC is applied. In this case the molecular HOMO is pinned at 
$E_\mathrm{F}$ and the transmission remains large. Interestingly this is a situation similar to the one we discussed 
in our previous work in the context of tight-binding Hamiltonian \cite{tbsic}. In this case the suppression of the 
conductance may further originate
from the derivative discontinuity of the potential, which may move the peak in the conductance away from $E_\mathrm{F}$. Since this
is not described by the ASIC scheme, we believe that the current in this case may be erroneously large. 

\begin{figure}[ht]
\begin{center}
\includegraphics[width=9.0cm,clip=true]{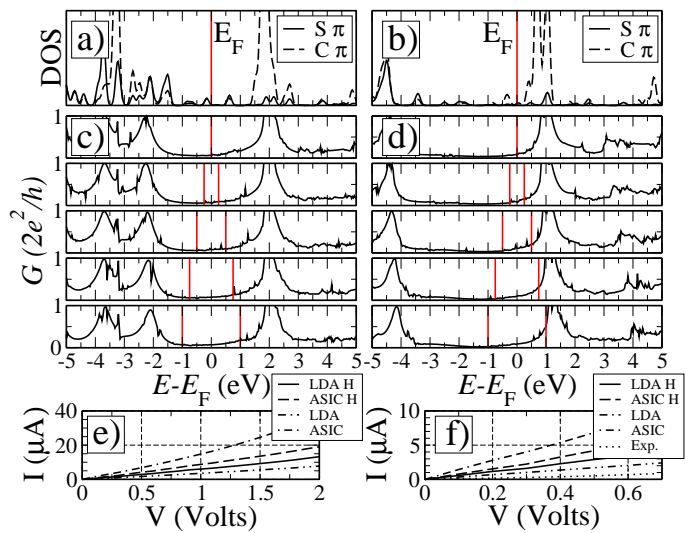}
\end{center}
\caption{\small{Transport properties of a BDT molecule attached to two hollow sites at both sides of the
junctions. In this case the H atoms of the thiol groups are left bounded to S and do not dissociate in
forming a thiolate end group. The left plots correspond to LDA and the right ones to ASIC. 
The upper panels are the DOS of the S and C $\pi$ orbitals ((a) and (b)), the middle are the transmission 
coefficients as a function of energy for various bias ((c) and (d)) and the lower are the $I$-$V$ curves 
for the system both with and without the hydrogen atoms attached to the sulfur. Figure (f) is a zoom of (e) 
and compares our results with experiments from reference \cite{Tao}. The vertical lines in (c) and (d) 
mark the bias window.}}
\label{Fig16}
\end{figure}

The results for the case where BDT is attached to the hollow sites but hydrogen atoms remain attached to thiol anchoring groups
are presented in figure \ref{Fig16}. The total energy for this situation is 1.47~eV higher than that obtained when
the end groups are thiolate and the remaining two H atoms form a H$_2$ molecule.
As can be seen from the DOS in panels (a) and (b) and the transmission coefficients in panels (c) and (d), the transport is 
now through the LUMO of the system, the energy of which is lowered only slightly by ASIC. Hence, the ASIC zero-bias 
conductance of 0.09~$G_0$ is higher than the value of 0.06~$G_0$ calculated with LDA. 
However it is important to note that the small, but significant, downshift of the LUMO state caused by ASIC is
only an artifact of our atomic approximation to SIC. As an empty state, the SIC for the LUMO is not defined and
no corrections should be applied. 

\subsection{Benzenedimethanethiol}

\begin{figure}[ht]
\begin{center}
\includegraphics[width=8.0cm,clip=true]{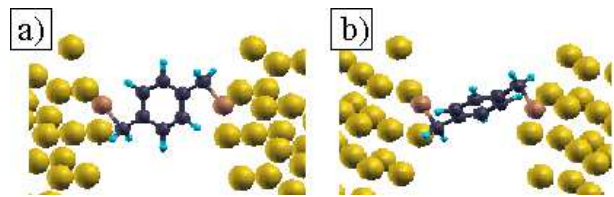}
\end{center}
\caption{\small{BDMT isomers attached to the hollow site of the Au (111) surface. 
The sulfur-surface distance is 1.9\AA. Color code: Au=yellow (or light gray), C=black, 
S=dark yellow (or gray), H=blue (or dark gray).}}
\label{Fig17}
\end{figure}

The second device studied consists of a benzenedimethanethiol (BDMT) molecule attached to two gold fcc (111) surfaces. 
Two different isomers of are investigated; in the first the sulfur atoms of the thiol groups lie in the same plane 
of the benzene ring (figure \ref{Fig17}a), while in the second they are aligned out of plane (figure \ref{Fig17}b). In both cases they
anchor the molecule to the hollow site of the gold fcc (111) surface.

\begin{figure}[ht]
\begin{center}
\includegraphics[width=9.0cm,clip=true]{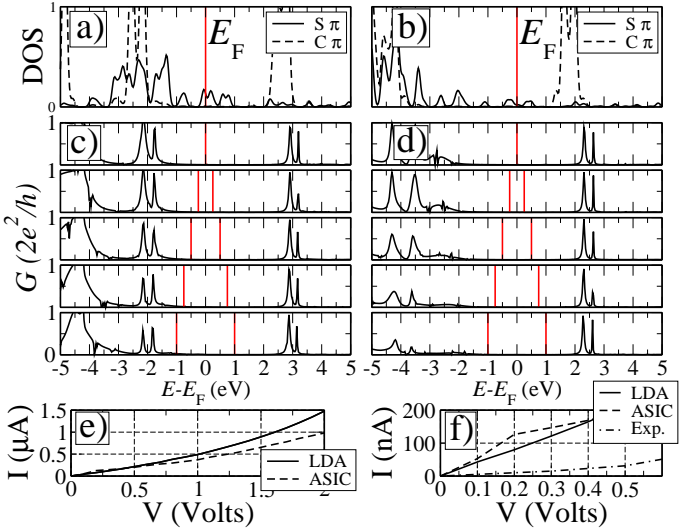}
\end{center}
\caption{\small{Transport properties of a BDMT molecule attached to the gold (111) hollow site for the isomer of figure
\ref{Fig17}. The left plots correspond to LDA results and the right ones to ASIC. The upper panels are the DOS of the S 
and C $\pi$ orbitals ((a) and (b)), the middle are the transmission coefficients as a function 
of energy for various bias ((c) and (d)) and the lower are the $I$-$V$ curves. Figure (f) is a zoom of
(e) and compares our results with experiments from reference \cite{Tao}. The vertical lines in
(c) and (d) mark the bias window.}}
\label{Fig18}
\end{figure}

The orbital resolved DOS, transmission coefficients and $I$-$V$ curves for the first isomer are presented
in figure \ref{Fig18} for both LDA and ASIC. In this case, as one can see from the orbital resolved density of states (figures 
\ref{Fig18}a and \ref{Fig18}b), the HOMO-LUMO gap is much larger than that of BDT. The ASIC again has the effect of shifting 
the occupied orbitals downwards in energy, but this has little effect on the transport since the resonances in the transmission coefficients 
due the HOMO and LUMO lie well outside the bias region investigated.

Therefore, ASIC increases the HOMO-LUMO gap but the actual transmission coefficient around $E_\mathrm{F}$
does not change much from its LDA value. The ASIC conductance at zero bias is calculated to be 0.004$G_0$, 
compared to 0.006$G_0$ obtained with LDA only. Both these values are one order of magnitude larger than the 
experimental value of  0.0006$G_0$ obtained by Xiao et. al. \cite{Tao}. It also follows that the $I$-$V$ curves 
(figures \ref{Fig18}e and \ref{Fig18}f) obtained with LDA and ASIC respectively are quite similar. 

\begin{figure}[ht]
\begin{center}
\includegraphics[width=9.0cm,clip=true]{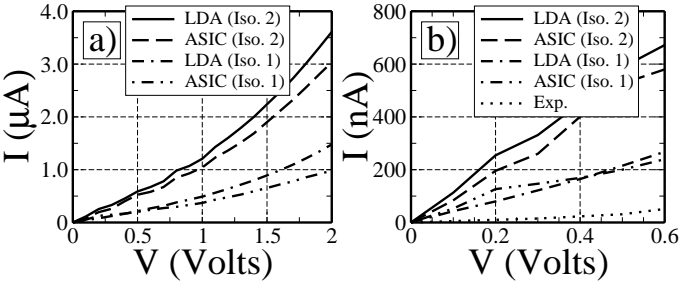}
\end{center}
\caption{\small{$I$-$V$ curves for both BDMT isomers attached to gold contacts. In both cases ASIC has only marginal
effects on the currents, which differ by a factor of about three for the two different isomers.}}
\label{Fig19}
\end{figure}

The $I$-$V$ curve calculated for the second isomer is presented in figure \ref{Fig19}. Also in this case the effects of
ASIC are minimal due to the large HOMO-LUMO gap. The conductance at zero bias is calculated to be 0.015$G_0$ 
with LDA, compared to the value of 0.013$G_0$ for ASIC. Interestingly this is about a factor 3 larger than 
that calculated for the first isomer.
%
 %

In general, in contrast to the case of BDT, for BDMT ASIC does not appear to improve the agreement between theory 
and experiments. The fact that the Fermi level of Au is pinned in the middle of the large HOMO-LUMO gap of BDMT, which is 
already well described by LDA, makes the ASIC corrections rather marginal to transport at moderate voltages. Given the fact that 
the geometry of the anchoring is not known and that we have only investigated one case, we may conclude that the 
disagreement between theory and experiments may simply lie in the details of the anchoring geometry. 

\subsection{Biphenyldithiol}

\begin{figure}[ht]
\begin{center}
\includegraphics[width=7.5cm,clip=true]{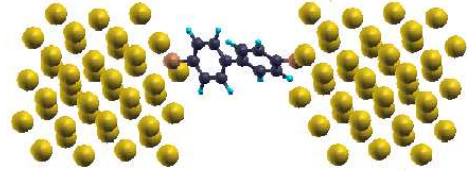}
\end{center}
\caption{\small{BPD molecule attached to the hollow site of the Au (111) surface. 
The sulfur-surface distance is 1.9\AA. Color code: Au=yellow (or light gray), C=black, 
S=dark yellow (or gray), H=blue (or dark gray).}}
\label{Fig20}
\end{figure}

The third and final system investigated is that of biphenyldithiol (BPD) attached to two gold surfaces as 
shown in figure \ref{Fig20}. Once again we consider only the gold fcc (111) hollow site as the anchoring site. 
The planes of the two benzene rings are usually tilted with respect to each other by an angle known as the torsion 
angle, that in this case is calculated to be 37$^o$.

\begin{figure}[ht]
\begin{center}
\includegraphics[width=9.0cm,clip=true]{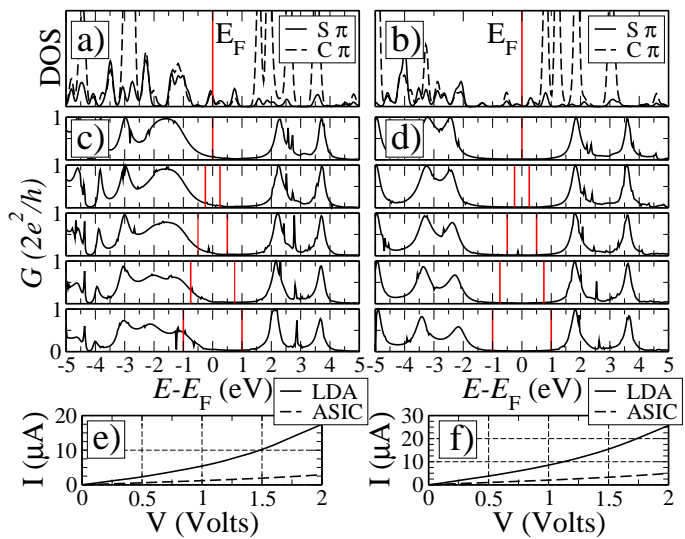}
\end{center}
\caption{\small{Transport properties of a BPD molecule attached to the gold (111) hollow site. The left
plots correspond to LDA and the right ones to ASIC. The upper panels are the DOS of the S 
and C $\pi$ orbitals ((a) and (b)), the middle are the transmission coefficients as a function 
of energy for various bias ((c) and (d)) and the lower are the $I$-$V$ curves. Figure (e) is the $I$-$V$ 
curve for a torsion angle of $37^{\circ}$ and (f) is for a torsion angle of $0^{\circ}$. The vertical lines in
(c) and (d) mark the bias window.}}
\label{Fig21}
\end{figure}

The orbital resolved DOS, transmission coefficients and $I$-$V$ curves for this system are presented
in figure \ref{Fig21} for both LDA and ASIC. Once again, ASIC has the effect of lowering the energy of the 
occupied molecular orbitals, as can be seen from the DOS (panels (a) and (b)). This in turn results in
opening up the conductance gap in the transmission as shown in panels (c) and (d) of figure \ref{Fig21}. 
For this molecule, the HOMO is 
near $E_\mathrm{F}$ as in the case of BDT, giving a LDA conductance of 0.07$G_0$ at zero bias. When ASIC 
is applied, the HOMO is shifted downwards in energy and outside the bias window for voltages up to 2~Volt. 
In this case the ASIC conductance drops to 0.018$G_0$ at $E_\mathrm{F}$ and the whole low bias $I$-$V$
curve shows rather smaller current with respect to its LDA counterpart.

The optimum torsion angle is $37^{\circ}$\cite{BPDtorang}. However, this may fluctuate  due to temperature or 
when the molecule is under strain in a breaking junction. Panel (e) of figure \ref{Fig21} shows the $I$-$V$ curves 
calculated for the equilibrium torsion angle of $37^{\circ}$, whereas panel (f) shows the result for the case when
the benzene rings are in the same plane (i.e. when the torsion angle is $0^{\circ}$). Reducing the 
torsion angle causes an increase in the transmission since the overlap between the $\pi$ orbitals is increased. 
The conductance at zero bias for a torsion angle of $0^{\circ}$ is 0.09$G_0$ with LDA only, and 0.024$G_0$ 
when ASIC is applied. 

These results show that ASIC has an effect on BPD similar to the one it has on BDT, i.e. it shifts the HOMO 
downwards in energy and reduces the zero-bias conductance. In this case however even our ASIC results differ from the
experimental data of Dadosh et. al. \cite{Dadosh} by several orders of magnitude. Since we are essentially in
a tunneling situation, the magnitude of the current is severely dependent on the tunneling matrix elements, which
in turn are rather sensitive to the details of the anchoring geometry. Although the hollow site is believed to be the
preferential site, such a large discrepancy between theory and experiments could be attributed to the contact
geometry.
 
\section{Conclusion}

In view of the many contradictory results for the transport properties of molecular junctions, 
first principles computational tools are becoming increasingly important. These, however, also suffer
from some uncertainty in the results, rooted in the approximations used for computing the junction
electronic structure. In this work, we have addressed the problem of the self-interaction error in DFT and
we demonstrate how a simple self-interaction correction scheme can be used to improve the 
agreement between theory and experiments. In particular, we have considered the problem of 
correctly reproducing the molecular ionization potential as the negative of the single particle
Kohn-Sham HOMO energy.

Our ASIC scheme is capable of achieving such a description, and for most of the molecules investigated 
this has the effect of moving conductance resonances away from the Fermi level of the electrodes. 
Thus, molecular junctions with zero bias resonant transport through the HOMO become insulating when 
calculated with ASIC. We then investigated transport through BDT attached to the gold (111) 
surfaces in several different anchoring geometries. In general, we have found a systematic reduction of
the low bias current when ASIC is applied. Most importantly, we have demonstrated that a number of
anchoring geometries all yield similar low bias conductivities, which in most cases are rather insensitive to
bond stretching or bond angle changes. This suggests that several anchoring geometries might contribute to 
the peaks in the typical experimental conductance histograms, thus providing a natural explanation for their
stability.

Nevertheless, even with ASIC, some problems still remain. In general, if the first molecular orbital to enter the bias window as the
bias increases is the LUMO, then our method potentially fails. In fact, ASIC only partially restores 
the DFT derivative discontinuity, and erroneously $\epsilon^\mathrm{KS}_\mathrm{HOMO}(N+\delta)\sim
\epsilon^\mathrm{KS}_\mathrm{LUMO}(N)$. Since the size of the derivative discontinuity is unknown,
it is difficult at this time to evaluate the impact of such an error. A second problem is connected to the 
accuracy of the calculation of the wave-functions. When
no molecular levels appear at the Fermi level, the transport is essentially tunneling like, and therefore
becomes crucially dependent on the accuracy of the calculated hopping integrals between the molecule
and the electrodes. ASIC does not drastically improve the quality of the LDA wave-function and thus
one might expect some errors in the tunneling transmission coefficients. Additionally, we do not correct
the surface atoms of the electrodes for SI, so it is expected that our ASIC results in general produce 
overextended orbitals and hence a systematically overestimated current. Finally, it is worth noting that 
ASIC still overestimates the polarizability of molecules \cite{ASICPOL, Kummel}, with a quantitatively 
incorrect prediction of the response exchange and correlation field. The use of XC potentials constructed 
from exact charge densities \cite{Baer} which correct both the molecule and the metallic surfaces may offer 
a solution to these problems. However, for some of the molecules investigated the disagreement between theory 
and experiments is still of several orders of magnitude. This is difficult to associate with any systematic 
error in the theory, and might be connected to the specific contact geometry. 

In conclusion, NEGF combined with the ASIC electronic structure method offers a description
of transport through molecules which is considerably improved with respect to standard NEGF LDA/GGA. 
Since this is also a computationally undemanding scheme, it can be employed for large systems,
and therefore can become an important tool for systematically predicting the performance of
molecular junctions. 

\section*{Acknowledgments}

We thank Alex~Reily Rocha, Chaitanya~Pemmaraju, Kieron~Burke and Alessio~Filippetti for useful discussions. 
This work is funded by Science Foundation of Ireland (grant SFI02/IN1/I175). Computational resources have 
been provided by the HEA IITAC project managed by the Trinity Center for High Performance Computing and by
ICHEC.
 \\

\end{document}